\documentclass[a4paper,12pt]{article}

\usepackage{graphicx} 
\usepackage{here}
\usepackage{setspace}
\usepackage{cite} 
\usepackage{url}  
\usepackage[latin1]{inputenc} 
\usepackage{textcomp} 
\usepackage{tabularx} 

\usepackage{sectsty}

\title{Effects of Growth on Dinitrogen on the Transcriptome and Predicted Proteome of \textit{Nostoc} PCC 7120}

\author{
R\"obbe W\"unschiers$^*$\\
\small{University of Cologne, Institute for Genetics, Z\"ulpicher Str.\ 47, D-50674 Cologne}
\and
Rikard Axelsson\\
\small{AstraZeneca AB, SE-151 85 S\"odert\"alje}
\and 
Peter Lindblad\\
\small{Uppsala University, Dep.\ of Physiol.\ Botany, EBC, Villav\"agen 6, SE-752 36 Uppsala}}

\date{}

\begin{document}

\maketitle

\begin{center}
\vspace*{2cm}
$^*$\textbf{Corresponding Author}\\robbe.wunschiers@uni-koeln.de\\
\end{center}

\newpage

\section*{Abstract}
Upon growth on dinitrogen, the filamentous cyanobacterium \textit{Nostoc} PCC 7120 initiates metabolic and morphological changes. We analyzed the expression of 1249 genes from major metabolic categories under nitrogen fixing and non-nitrogen fixing growth. The expression data were correlated with potential target secondary structures, probe GC-content, predicted operon structures, and nitrogen content of gene products. Of the selected genes, 494 show a more than 2-fold difference in the two conditions analyzed. Under nitrogen-fixing conditions 465 genes, mainly involved in energy metabolism, photosynthesis, respiration and nitrogen-fixation, were found to be stronger expressed, whereas 29 genes showed a stronger expression under non-nitrogen fixing conditions. Analysis of the nitrogen content of regulated genes shows that \textit{Nostoc} PCC 7120 growing on dinitrogen is freed from any constraints to save nitrogen. For the first time the expression of high light-induced stress proteins (HLIP-family) is shown to be linked to the nitrogen availability.

\newpage

\section*{Introduction}
Cyanobacteria are intensively used in biotechnology. We are interested in growing cyanobacteria in photobioreactors with the goal of hydrogen production and biomass generation \cite{Wuenschiers2003a,Dutta2005}. Hydrogen metabolism is closely linked to dinitrogen assimilation in nitrogen-fixing bacteria. Heterocystous nitrogen-fixing cyanobacteria like \textit{Nostoc} sp.\ strain PCC 7120 (\textit{Nostoc} PCC 7120) respond to the deprivation of combined nitrogen with morphological changes, i.e., the formation of heterocysts. These specialized non-dividing cells develop more or less equidistantly along the filaments with a ratio of about one heterocyst per 10 vegetative cells. In comparison to vegetative cells, the heterocyst is larger and more rounded. It provides an environment with low oxygen partial pressure since it lacks oxygen-evolving photosystem II activity and has a higher respiration rate \cite{Boehme1998}. Furthermore, it is surrounded by a thick glycolipidic cell wall that reduces the diffusion of oxygen. Heterocysts provide amino acids and receive carbohydrates from their neighboring vegetative cells. The differentiation process begins within a few hours after combined-nitrogen deprivation, i.e. growth on dinitrogen as sole nitrogen source, and requires approximately 24 hours to complete. Although the detailed cellular processes underlying heterocyst formation and its regulation are not well understood, many single components involved in this process have been described (for reviews see: \cite{Golden1998,Adams2000,Wolk2000,Meeks2002a,Herrero2004}). A key protein in heterocyst formation is NtcA, 
which is the global nitrogen regulator that controls, e.g., the expression of genes essential for heterocyst development such as the ABC-type transporter genes \textit{devABC}
and the regulator genes \textit{hetR}
and \textit{patS}.
Among many other factors that have been demonstrated to be important for heterocyst development are enzymes that are involved in the formation of heterocyst-specific glycolipids and polysaccharides.
The driving forces for these structural changes occurring during heterocyst formation are metabolic requirements: the dinitrogen fixing multi-enzyme complex nitrogenase requires a low oxygen partial pressure and a high supply of ATP and reduction equivalents \cite{Dixon2004}. ATP may be generated by either cyclic photophosphorylation or oxidative phosphorylation while low-potential electrons may be generated from the degradation of carbohydrates produced during photosynthesis \cite{Haselkorn1992}. Obviously, profound regulatory events coincide with growth on dinitrogen.

A powerful tool to study gene expression and its regulation is the DNA-microarray technique. Until now only few approaches about DNA-microarray based gene-expression analyses with heterocystous cyanobacteria have been reported. With respect to nitrogen metabolism, only one experiment has been described \cite{Ehira2003,Sato2004}. The authors used a segment-based DNA-microarray, where each segment covers up to 8 predicted genes \cite{Ehira2003}. This experimental setup does not allow expression analysis of individual genes. 

Previously, we analyzed the expression of individual genes and operons of \textit{Nostoc} spp.\ that are involved in nitrogen metabolism (for a review see: \cite{Wuenschiers2003a}). Here we describe an oligo-nucleotide based DNA-microarray expression analysis, where each gene is covered by up to 10 unique probes. 
We employed a novel, recently developed microarray technique where probe synthesis, hybridization, and signal detection take place in one device at strongly controlled physical conditions \cite{Baum2003,Gueimil2003}. 
The expression data were used to a) validate the technique employed and b) obtain a global overview about the effect of growth on dinitrogen on \textit{Nostoc} PCC 7120.
In addition, we describe a convenient data-processing pipeline based on a MySQL database and a web-based graphical user interface. This front-end allows users to overlay gene-expression data on KEGG (Kyoto Encyclopedia of Genes and Genomes) pathway maps. 

\section*{Material and Methods}

\subsection*{Strains and Culture Conditions}
The cyanobacterium \textit{Nostoc} sp.\ strain PCC 7120 (formerly \textit{Anabaena} sp.\ strain PCC 7120) was grown on either dinitrogen (nitrogen fixing) or combined nitrogen (non-nitrogen fixing) in batch cultures. Non-nitrogen fixing conditions were obtained by growing cells in BG11$_{0}$ supplemented with 5 mM NH$_{4}$Cl and 10 mM HEPES. Nitrogen-fixing conditions were obtained by growing cells in BG11$_{0}$. All cultures were grown and harvested as previously described \cite{Oxelfelt1995,Tamagnini1997}. 

\subsection*{Preparation of Biotin Labeled, Fragmented cRNA}
Isolation of total RNA from \textit{Nostoc} PCC 7120 was performed as previously described \cite{Axelsson2002}. From 10 $\mu$g of total RNA, low molecular weight RNA, e.g., tRNA and 5S rRNA, were removed by size exclusion chromatography (MEGAclear kit, Ambion). To remove 16S and 23S rRNA, the MICROBExpress kit from Ambion was used. The remaining RNA was linearly amplified by a modified Eberwine protocol \cite{Eberwine1992} as follows. If not differently stated, all enzymes and chemicals were purchased from Invitrogene.
\textit{First strand synthesis}: The pelleted RNA from the previous mRNA-enrichment steps was resuspended in 4.25 $\mu$l water and mixed with 1 $\mu$l of T7 random hexamers (0.5 $\mu$g/$\mu$l; 5'-GGC CAG TGA ATT GTA ATA CGA CTC ACT ATA GGG AGG CGG NNN NNN-3'). Following incubation at 70\textcelsius\ for 10 min, 4\textcelsius\ for 2 min and 23\textcelsius\ for 5 min, 3.75 $\mu$l reaction mix (2 $\mu$l 5x first strand synthesis buffer (1 $\mu$l 0.1 M DTT, 0.5 $\mu$l 10 mM dNTP mix, 0.25 $\mu$l 40 U RNase OUT) and 200 U Superscript II polymerase) was added to the RNA/primer mix. First strand synthesis reaction was performed with the following temperature scheme: 37\textcelsius\ for 20 min, 42\textcelsius\ for 20 min, 50\textcelsius\ for 15 min, 55\textcelsius\ for 10 min and 65\textcelsius\ for 15 min. After adding 0.5 $\mu$l RNase H the reaction mix was incubated for another 30 min at 37\textcelsius\ and 2 min at 95\textcelsius. 
\textit{Second strand synthesis}: The product of the first strand synthesis was mixed with 43.8 $\mu$l water and 15 $\mu$l 5x second strand synthesis buffer (20 U DNA-polymerase I, 1.5 $\mu$l 10 mM dNTP and 1 U RnaseH) and incubated for 2 h at 16\textcelsius. After addition of 10 U T4 DNA-polymerase the reaction mix was first incubated at 16\textcelsius\ for 15 min and then at 70\textcelsius\ for 10 min. 
\textit{Isolation of ds-cDNA}: Double stranded cDNA was isolated from the product of second strand synthesis according to standard procedures \cite{Maniatis1982}. 
\textit{In vitro transcription}: The pelleted ds-cDNA was resuspended in 1.5 $\mu$l water. The MEGAscript T7 kit (Ambion) was used for \textit{in vitro} transcription. In addition to the standard nucleotides, 3.75 $\mu$l 10 mM Bio-16-CTP (NEN) and 3.75 $\mu$l 75 mM Bio-11-UTP (Roche) were added to the reaction mix. This led to the formation of biotinylated cRNA.
\textit{cRNA-isolation}: The RNeasy kit (Qiagen) was applied for cRNA-isolation. All steps were performed according to the manufacturer's instructions. 
\textit{cRNA-fragmentation}: For cRNA-fragmentation 15 $\mu$g cRNA was resuspended in 2.5 $\mu$l water and 2.5 $\mu$l 2x fragmentation buffer (5x stock: 200 mM Tris, 150 mM Mg-acetate, 500 mM K-acetate, pH 8.1). The reaction mix was incubated for 5 min at 94\textcelsius. The fragmentation reaction was performed immediately prior to hybridization. 

\subsection*{Oligonucleotide Probe Selection}
A unique \textit{Nostoc} PCC 7120 probe set (as many 25-mer probes per open reading frame (ORF) as possible) was calculated based on the full genome sequence (retrieved online from CyanoBase: \url{http://www.kazusa.or.jp/cyanobase/Anabaena/index.html}) using a combination of sequence uniqueness criteria and rules for selection of oligonucleotides likely to hybridize with high specificity and sensitivity. The selection criteria were as described in Lockhart \textit{et al}.\ \cite{Lockhart1996} with modifications for the longer probes used here (25-mers instead of 20-mers). If available, 10 unique probes per ORF were used in the experiments.

\subsection*{DNA-Microarray Production and \textit{In Situ} Oligonucleotide Synthesis}
Light-activated \textit{in situ} oligonucleotide synthesis was performed as described by Singh-Gasson \textit{et al}.\  \cite{Singh-Gasson1999} using a digital micromirror device, which is part of the geniom one device (febit biotec GmbH, Heidelberg/Germany). The synthesis was performed within the geniom one device on an activated three-dimensional reaction carrier consisting of a glass-silica-glass sandwich (DNA-processor). Four individually accessible microchannels (referred to as arrays), etched into the silica layer of the DNA-processor, were connected to the microfluidic system of the geniom device. Using standard DNA-synthesis reagents and 3'-phosphoramidites with a photolabile protecting group \cite{Beier2000,Hasan1997}, oligonucleotides were synthesized in parallel in all four translucent arrays of one reaction carrier. Prior to synthesis, the glass surface was activated by coating with a silane-bound spacer.

\subsection*{Hybridization}
Non-competitive hybridizations were performed with 7.5 $\mu$g fragmented cRNA (see above) in a final volume of 10 $\mu$l. The hybridization solution contained 100 mM MES (pH 6.6), 0.9 M NaCl, 20 mM EDTA, 0.01\% (v/v) Tween 20, 0.1 mg/ml sonicated herring sperm DNA, and 0.5 mg/ml BSA. RNA-samples were heated in the hybridization solution to 95\textcelsius\ for 3 min followed by 45\textcelsius\ for 3 min before being placed in an array which had been prehybridized for 15 min with 1\% (w/v) BSA in hybridization solution at room temperature. Hybridizations were carried out at 45\textcelsius\ for 16 h. After removing the hybridization solutions, arrays were first washed with non-stringent buffer (0.005\% (v/v) Triton X-100 in 6x SSPE) for 20 min at 25\textcelsius\ and subsequently with stringent buffer (0.005\% (v/v) Triton X-100 in 0.5x SSPE) for 20 min at 45\textcelsius. After washing, the hybridized RNA was fluorescence-stained by incubating with 10 $\mu$g/ml streptavidin-phycoerythrin and 2 $\mu$g/$\mu$l BSA in 6x SSPE at 25\textcelsius\ for 15 min. Unbound streptavidin-phycoerythrin was removed by washing with non-stringent buffer for 20 min at 25\textcelsius.

\subsection*{Detection and Data Processing}
The CCD-camera based fluorescence detection system, equipped with a Cy3 filter set, integrated into the geniom one automate was used. 36 pixels per spot were available for data analysis. Processing of raw data, including background correction, array to array normalization and determination of gene-expression levels, as well as calculation of expression differences were performed as described before \cite{Zhou2002}. All steps were carried out using the PROP-algorithm of the geniom application software which is based on the MOID-algorithm described by Zhou and Abagyan \cite{Zhou2002}. Background correction is based on probes with no corresponding mRNA-target and the average of the lowest 5\% expressed genes. Data normalization is based on iteratively correcting the raw data on non-regulated genes. It could be shown previously in a comparative study of \textit{Saccharomyces cerevisiae} gene expression with three independent techniques (i.e., Affymetrix GeneChips, geniom one microarrays, and cDNA microarrays) that expression differences greater than \textpm1.5-times are significant for the geniom one platform applied here \cite{Baum2003}. In our study we extend this range such that the upper or lower bound had to be greater than \textpm2.

\subsection*{Operon Analysis}
For annotating ORFs belonging to operons, the assignment by Ermolaeva \textit{et al}.\ \cite{Ermolaeva2001} was used. Only ORFs belonging to operons with probabilities above 60\% were included in the analysis. For 51 ORF-pairs the maximum gene-expression difference was calculated (operon set). The same number of ORF-pairs was set up by randomly choosing ORFs that do not belong to the same operon (non-operon set). For statistical analysis, data from four arrays were pooled, yielding 204 data points per set, and subjected to non-parametric significance tests (Mann-Whitney- and Kolmogorov-Smirnov test).

\subsection*{Probe Secondary Structure and GC-Content Analysis}
In order to correlate probe secondary structures and GC-contents with their expression value, the secondary structure of each ORF was calculated using the Vienna RNA Package Version 1.4 \cite{Hofacker1994}. The probes were aligned to their corresponding ORF and the potential probe structure extracted. The number of hybridized (stem duplexes) versus free (loops and bulges) probe nucleotides as well as the probe's GC-content were used in further analysis.

\subsection*{HyDaBa Database}
All gene-expression data obtained are saved in the Hydrogenase Database (HyDaBa, \url{http://hydaba.uni-koeln.de}). This relational database allows cross-linking of the expression data with the annotated genome data from NCBI (\url{http://www.ncbi.nlm.nih.gov}) and Cyano\-base (\url{http://www.kazusa.or.jp/cyano/Anabaena/cgi-bin/category\_ana.cgi})\linebreak and pathway maps available from KEGG (http://www.genome.jp/kegg). The latter is achieved in real-time via a SOAP-interface. HyDaBa is based on a Apache Webserver (\url{http://www.apache.org}), MySQL database (\url{http://www.mysql.com}) and a front-end programmed in PHP (\url{http://www.php.net}). All data are publicly accessible via this web interface.

\section*{Results and Discussion}
In the present study we analyzed the expression of 1249 selected genes from 16 metabolic categories (ca.\ 20\% of the complete genome) of \textit{Nostoc} PCC 7120 cultures under nitrogen fixing and non-nitrogen fixing conditions (Tab.\ 1). Therefore we applied a DNA-microarray based approach (Fig.\ 1).

\subsection*{Preparation of the DNA-Processor}
Oligonucleotide synthesis, hybridization with target cRNA, and signal detection were performed with one single device, named geniom one (febit biotech GmbH, Heidelberg/Germany, see \cite{Baum2003}). 25-mer oligonucleotide probes were synthesized \textit{in situ} on the DNA-microarray surface. In order to obtain a broad picture of gene-expression differences between nitrogen fixing and non-nitrogen fixing \textit{Nostoc} PCC 7120 cultures, 500 manually and 749 randomly selected target genes from all major metabolic categories were analyzed (Tab.\ 1; \cite{Kaneko2001}). This selection was based on the genome sequence and annotation available from the CyanoBase consortium (\url{http://www.kazusa.or.jp/cyanobase/Anabaena/index.html}). In order to ensure reproducibility of the microarray analysis, up to 10 unique 25-mer oligonucleotide probes per target ORF were distributed randomly over the DNA-processor. Due to their small size, 132 ORFs were represented by less than 10 unique probes. Of theses, 78 represent unknown and 15 hypothetical proteins, respectively. Of the remaining only 5 ORFs were represented by less than 4 unique probes. Among those is the ORF encoding the heterocyst differentiation related protein PatN (alr4812). 

\subsection*{General Data Analysis}
Figure 2 shows a section of four arrays used in this analysis. Due to \textit{in situ} probe synthesis with a digital micromirror device both the spot morphology and topology are extremely homogeneous. The use of one physical surface for all arrays and the fixed placement of the slide during all processing steps results in very low experimental variation. In order to visualize the signal-to-noise ratio the fluorescence-intensity ratio of either two RNA-samples from different growth conditions or from two RNA-samples from the same growth condition are plotted in double-logarithmic scales (Fig.\ 3). It can be clearly seen that the comparison of two different metabolic states scatters much broader than the self-to-self comparison. The variance of the data is displayed by their respective Pearson correlation value r$^{2}$. Five ORFs in the self-to-self comparison show an unexpected large variation. In three cases (asr7152, alr7535 and alr7580) this can be explained by their low, close to threshold fluorescence-signal intensity.

\subsection*{Operon Analysis}
In order to obtain a different insight into the quality of the obtained expression data, we exploit the fact that the majority of genes within an operon should exhibit similar expression levels. Therefore, we compared the expression-level differences of ORF-pairs that either belong to one single operon or not (Fig.\ 4). The grouping of ORF-pairs into either the operon set or the non-operon set was based on data published by Ermolaeva \textit{et al}.\  \cite{Ermolaeva2001}. Their main criterion to assign operons is conserved ORF-neighborhood over a broad range of microbes. As expected, the expression difference between ORFs not belonging to the same operon are significant larger (p$<$0.001) than between ORF-pairs belonging to one single operon.

\subsection*{GC-Content and Probe Secondary Structure Analysis}
It is often observed in oligonucleotide-based DNA-microarray experiments that probes directed against one single transcript show large hybridization level variations. The reason for this fluctuation remains still unknown and is one major reason for the necessity to calculate the expression value for each transcript from several unique probes. We analyzed the influence of both GC-content and secondary structure formations on the hybridization signal.

Figure 5A shows the correlation between the number of guanine and cytosine nucleotides (GC-content) in the 25-mere probes and the corresponding hybridization signal. There is no probe with less than 3 or more than 17 GCs. In the range between 7 and 13 GCs there is a clear linear correlation between GC-content and hybridization signals. To many (more than 50\%) or to few (less than 25\%) GCs in the probe result in non-linear behavior. As can be seen from the number of observations given in the plot, less than 0.7\% (328 out of 48208) of all unique probes are affected by this non-linear behavior. At the current stage it is not possible to draw conclusions from the extremes on both sides of Figure 5A because they are only represented by few data. In contrary to the GC-content, the predicted secondary structure (stems, loops and bulges) of the transcript has only little influence on the hybridization signal (Fig.\ 5B). This was expected because (a) the target was chopped into smaller fragments prior to hybridization, and (b) the hybridization conditions are set such that no secondary structures should form in either the probe or the target. 

The effect of individual probe hybridization signals is usually ignored in oligonucleotide-based DNA-microarray experiments. Instead, the average over all probes is used for each transcript. However, these effects have to be taken into account if only few probes are available for particular transcripts. Furthermore, we can conclude that a large portion of the hybridization signal variation is intrinsic to the probe sequence and can not be explained by currently known DNA-duplex formation physico-chemistry.

\subsection*{Data Processing and Visualization}
DNA-microarray experiments involve accumulation and management of large amounts of data. Apart from the experimental data, information from open access knowledge databases and sequence analysis are collected. To provide optimal accessibility to all data we set up a MySQL database on an Apache driven Internet server. The database holds both raw and processed data. Besides data management the database allows cross-connectivity of expresseion data with annotations from NCBI database, Cyanobase and KEGG (Fig.\ 6). In order to access and query the database (which has been coined HyDaBa) a PHP-based and Internet-accessible front-end has been developed. This front-end helps to query the data, guides the user to define and store new queries, allows data up- and download, and can be easily extended due to its modular setup with template pages. The most important feature of HyDaBa constitutes the mapping of gene-expression data onto metabolic charts from the KEGG database (Fig.\ 6). Technically, this has been achieved by using a SOAP interface \cite{Kawashima2003}. Equally important is the possibility to query for all data available for a given ORF. HyDaBa can be accessed at \url{http://hydaba.uni-koeln.de}. 

\subsection*{Global Differences in Gene Expression upon Growth on Dinitrogen}
Growth on dinitrogen as sole nitrogen source acts like a positive transcriptional switch in \textit{Nostoc} PCC 7120. There is a much larger fraction of genes stronger expressed under nitrogen fixing than under non-nitrogen fixing conditions and only a minority of genes shows a decreased expression level. Only 17 annotated and 12 hypothetical ORFs exhibit a significant higher expression under non-nitrogen fixing conditions (Tab.\ 2) whereas 281 annotated and 184 hypothetical ORFs are stronger expressed under nitrogen fixing conditions. In Figure 7, these gene-expression differences are clustered according to the participation of the corresponding ORF in specific metabolic categories. The strongest expressed genes participate in photosynthesis and respiration (K). Closer analysis reveals that 21 of the 29 strongest expressed genes in this group belong to photosynthesis, 11 of which are structural proteins of phycobilisomes (Tab.\ 3). These findings clearly illustrate the extensive energy demand for nitrogen fixation. Obviously, the cell expands its light harvesting complexes in order to direct more light energy to the photosystems and produce both more ATP and NADPH. The stronger expression of proteins involved in respiration underlines previous findings that the respiration rate is increased under nitrogen fixing conditions in cyanobacteria. It is believed that this process supports the removal of oxygen which otherwise would inactivate the nitrogenase enzyme complex.

\subsection*{Nitrogen Content of Gene Products}
It has been shown previously for yeast and \textit{E.\ coli} that enzymes involved in the assimilation of sulfur, nitrogen, or carbon are depleted in that respective atom \cite{Baudouin-Cornu2001}. Furthermore, upon sulfur depletion, yeast spares sulfur rich proteins by either reducing their expression level or by expressing isoforms that are poor in sulfur \cite{Fauchon2002}. We are interested in the global responds of \textit{Nostoc} PCC 7120 to the depletion of both inorganic and organic nitrogen in the culture medium. To address this question we tested if cells grown on dinitrogen as sole nitrogen source tend to spare nitrogen. This should be reflected in the nitrogen content of the gene products. As a rough approximation we assume that the expression level for a particular gene is proportional to the amount of protein the gene is encoding for. Then we can analyze whether the proteome shows differences with respect to the nitrogen content upon growth on dinitrogen as nitrogen source. Figure 8 shows the difference of expressed nitrogen atoms (number of nitrogen atoms per transcript times expression value) under nitrogen fixing and non-nitrogen fixing conditions. We observed a significant difference between these two conditions. Obviously, \textit{Nostoc} PCC 7120 does not save but spend nitrogen under nitrogen fixing conditions (Fig.\ 8). We conclude that the cells are freed from any limitations of nitrogen utilization once the nitrogenase enzyme complex is expressed and active. This has major implication for the utilization of cyanobacteria for, e.g., hydrogen gas production \cite{Dutta2005}. Although depletion of inorganic nitrogen is prerequisite for nitrogenase based production of hydrogen gas, the cells are not facing stress from nitrogen limitation. Thus, the biomass obtained from such photobioreactors will yield a rich food source for, e.g., livestock farming.

\subsection*{Heterocyst Related Genes}
As a key global regulator, NtcA plays an important role in the expression of many genes involved in heterocyst differentiation and nitrogen assimilation. For the unicellular, non-differentiating cyanobacterium \textit{Synechococcus} PCC 7942 it has been shown that the binding affinity of NtcA to its target DNA-sequence is elevated by 2-oxoglutarate \cite{Vazquez-Bermudez2003,Vazquez-Bermudez2002}. Thus, 2-oxoglutarate exerts a direct role on NtcA-mediated transcription activation. Furthermore, it plays a central role in sensing the nitrogen status, or rather the C/N-balance and acts as a substrate in glutamate synthesis, which in turn is one of the first metabolic steps of ammonium assimilation (Fig.\ 9). We found key enzymes catalyzing the synthesis of 2-oxoglutarate, aconitase hydratase (2.8-times; EC 4.2.1.3) and isocitrate dehydrogenase (4.7-times; EC 1.1.1.42), respectively, being stronger expressed under nitrogen fixing condition (Fig.\ 9). The \textit{hetC} gene (alr2817), which encodes a putative ABC-transporter that is essential for heterocyst formation, has been shown to be a direct target of the transcriptional regulator NtcA \cite{Muro-Pastor1999}. Indeed we see a 5.8-time stronger expression under nitrogen fixing conditions. Table 4 shows expression differences for all known ORFs involved in heterocyst formation included in this study.

\subsection*{Nitrogen Metabolism Related Genes}
The conversion of dinitrogen to ammonia, catalyzed by the nitrogenase enzyme complex, is only the first step in a series of reactions that make nitrogen available to the cell. The nitrogenase enzyme complex provides two products that are metabolized, hydrogen gas and ammonia, respectively. The former is taken up by an uptake hydrogenase while the latter is incorporated to glutamate by the glutamine synthase yielding glutamine. Figure 9 gives an overview over the main pathways and enzyme complexes involved in nitrogen fixation. The nitrogenase consists of three subunits, the molybdenum-iron protein alpha chain (NifD, all1454), the molybdenum-iron protein beta chain (NifK, all1440), and the iron protein (NifH, all1455). We found \textit{nifK} and \textit{nifH} to be more than 10-times stronger expressed under nitrogen fixing conditions. One ORF (\textit{nifH2}, alr0874) of yet unknown function that is paralogous (92\% similarity, 86\% identity) to the iron protein (\textit{nifH}, all1455) was found to be expressed at a very lower level and is only slightly stronger expressed under nitrogen fixing conditions (Fig.\ 9). Thus, the gene product of \textit{nifH2} is probably not involved in the nitrogen fixation reaction. 
The glutamine synthase (glutamate-ammonia ligase) is encoded by \textit{glnA} (alr2328). Northern blot studies in \textit{Nostoc} PCC 7120 have shown that the \textit{glnA} transcript is present in both nitrogen-fixing as well as non-nitrogen fixing cultures, but more abundant in the latter \cite{Tumer1983}. This is in accordance with our results.

\subsection*{Other Genes}
\subsubsection*{Phycobilisomes}
Phycobilisomes are the major light-harvesting complexes of cyanobacteria. These are multiprotein assemblies that are functionally associated with photosystem II and constitute up to 50\% of the total cellular protein. It has been shown previously that phycobilisomes serve as a nitrogen storage. Upon nitrogen starvation they can be completely degraded within two days. Phycobilisome degradation is thought to provide substrates for protein biosynthesis. As discussed above, we observed strong expression of major components involved in photosynthesis under nitrogen fixing conditions. Accordingly, all proteins included in this analysis and that constitute phycobilisomes are around 3-times stronger expressed under nitrogen fixing conditions.

\subsubsection*{Respiratory Terminal Oxidases}
\textit{Nostoc} PCC 7120 possesses three cytochrome c oxidase gene clusters, cox1 (alr0950, alr0951, alr0952), cox2 (alr2514, alr2515, alr2516), and cox3 (alr2729, alr2730, alr2731, alr2732, alr2734), respectively \cite{Valladares2003}. While cox1 and cox2 are homologous to aa$_3$-type cytochrome c oxidases \cite{Jones2002,Schmetterer1994}, cox3 is most similar to alternative respiratory terminal oxidases \cite{Valladares2003}. The expression of cox2 and cox3 has been reported to be restricted to heterocysts \cite{Valladares2003}. In accordance to this result we see no difference in the expression of cox1, while cox2 and cox3 are more than 4-times stronger expressed under nitrogen fixing conditions (see Table 5).

\subsubsection*{Adaptations and Atypical Conditions}
One group of genes that is up to 11-times stronger expressed under non-nitrogen fixing conditions belong to the high light-induced proteins (HLIP-family; Tab.\ 6) \cite{Heddad2002,Montane2000}. These proteins belong to the CAB/ELIP/HLIP-superfamily and are evolutionary related to each other \cite{Heddad2002}. While CAB (chlorophyll a/b-binding) proteins are major constitutions of the light harvesting complexes, ELIPs (early light-induced proteins) and HLIPs are taking over photo-protective functions. The HLIP-family in pro- and eukaryotic photosynthetic organisms consists of more than 100 different stress proteins which have one membrane spanning alpha helix. They accumulate only transiently in photosynthetic membranes in response to light stress and have photoprotective functions. At the amino acid level, members of the HLIP-family are closely related to light-harvesting chlorophyll a/b-binding antenna proteins of photosystem I and II, present in higher plants and some algae. Despite this similarity it is believed that HLIP-proteins fulfill their photoprotective role by either transient binding of free chlorophyll molecules or by participating in energy dissipation \cite{Montane2000}. Photooxidative stress in not necessarily connected to high light fluxes but can also be caused by nutrient deprivation that ultimately lead to oversaturation of the photosynthetic electron transport chain. At this point one can only speculate why the HLIP-family is stronger expressed under non-nitrogen fixing conditions. Since the same light and temperature settings were employed for both growing conditions, one argument would be that more light is required (and thus ''consumed'') under nitrogen fixing conditions. Indeed the nitrogenase enzyme complex has immense energy and reducing power demands. In concert with the higher availability of nitrogen upon growth on dinitrogen (see above) we conclude that this growth condition frees \textit{Nostoc} PCC 7120 from stress. To our knowledge this is the first report about differential expression of members of the HLIP-family in cyanobacteria upon combined-nitrogen deprivation. Analysis of the location of the HLIP-family members shown in Table 6 reveals no link to known ORFs that are involved in nitrogen metabolism. Only asl0449 is located immediately downstream of the allophycocyanin alpha subunit (\textit{apcA}, all0450) and thus demonstrates its potential functional relation to photosynthesis. Future work might help to uncover the function of these proteins in pro- and eukaroytes.

\section*{Acknowledgements}
The authors like to thank Dr.\ M.\ Baum and N.\ Rittner from febit for technical assistance with the microarray experiments and H.\ Eckes for initial work on the database front-end. RW likes to thank Prof.\ D.\ Tautz for his continuous support. RW's work is part of the BMBF-funded Cologne University Bioinformatics Center (CUBIC). This study was financially supported by grants to PL from the Swedish Research Council, the Swedish Energy Agency, and the EU/NEST Project SOLAR-H (contract 516510).

\newpage

  \bibliography{/home/rw/Eigene/ReferencesJabRef/referencesjabref} 

\begin{thebibliography}{10}

\bibitem{Wuenschiers2003a}
W\"unschiers R., Lindblad P.. {\it Handbook of Photochemistry and
  Photobiology}ch.~Light-Dependent Hydrogen Metabolism and Generation by
  Cyanobacteria, :295-328.
\newblock American Scientific Publishers, North Lewis Way, CA, USA 2003.

\bibitem{Dutta2005}
Dutta D., De~D., Chaudhuri S., Bhattacharya S.~K.. Hydrogen production by
  {C}yanobacteria.  {\it Microb. Cell Fact.. } 2005;4:36.

\bibitem{Boehme1998}
B\"ohme H.. Regulation of nitrogen fixation in heterocyst-forming
  cyanobacteria.  {\it Trends in Plant Science. } 1998;3:346-351.

\bibitem{Golden1998}
Golden J.~W., Yoon H.~S.. Heterocyst formation in \textit{{A}nabaena}.  {\it
  Curr. Opin. Microbiol.. } 1998;1:623-629.

\bibitem{Adams2000}
Adams D.~G.. Heterocyst formation in cyanobacteria.  {\it Curr. Opin.
  Microbiol.. } 2000;3:618-624.

\bibitem{Wolk2000}
Wolk C.~P.. {\it Prokaryotic Development.}ch.~Heterocyst formation in
  \textit{Anabaena}., :83-104.
\newblock American Society for Microbiology, USA 2000.

\bibitem{Meeks2002a}
Meeks J.~C., Elhai J.. Regulation of cellular differentiation in filamentous
  cyanobacteria in free-living and plant-associated symbiotic growth states.
  {\it Microbiol. Mol. Biol. Rev.. } 2002;66:94-121.

\bibitem{Herrero2004}
Herrero A., Muro-Pastor A.~M., Valladares A., Flores E.. Cellular
  differentiation and the {N}tc{A} transcription factor in filamentous
  cyanobacteria.  {\it FEMS Microbiol. Rev.. } 2004;28:469-487.

\bibitem{Dixon2004}
Dixon R., Kahn D.. Genetic regulation of biological nitrogen fixation.  {\it
  Nat. Rev. Microbiol.. } 2004;2:621-631.

\bibitem{Haselkorn1992}
Haselkorn R., Buikema J.~W.. {\it Biological Nitrogen Fixation.}ch.~Nitrogen
  fixation in cyanobacteria., :166-190.
\newblock Chapman \& Hall, Inc. NY, USA 1992.

\bibitem{Ehira2003}
Ehira S., Ohmori M., Sato N.. Genome-wide expression analysis of the responses
  to nitrogen deprivation in the heterocyst-forming cyanobacterium
  \textit{{A}nabaena} sp. strain {PCC} 7120.  {\it DNA Res.. } 2003;10:97-113.

\bibitem{Sato2004}
Sato N., Ohmori M., Ikeuchi M., et al. Use of segment-based microarray in the
  analysis of global gene expression in response to various environmental
  stresses in the cyanobacterium \textit{{A}nabaena} sp. {PCC} 7120.  {\it J.
  Gen. Appl. Microbiol.. } 2004;50:1-8.

\bibitem{Baum2003}
Baum M., Bielau S., Rittner N., et al. Validation of a novel, fully integrated
  and flexible microarray benchtop facility for gene expression profiling.
  {\it Nucleic Acids Res.. } 2003;31:e151.

\bibitem{Gueimil2003}
G\"uimil R., Beier M., Scheffler M., et al. Geniom technology - the benchtop
  array facility.  {\it Nucleosides Nucleotides Nucleic Acids. }
  2003;22:1721-1723.

\bibitem{Oxelfelt1995}
Oxelfelt F., Tamagnini P., Salema R., Lindblad P.. Hydrogen uptake in
  \textit{{N}ostoc} strain {P}{C}{C} 73102: Effects of nickel, hydrogen, carbon
  and nitrogen.  {\it Plant. Physiol. Biochem.. } 1995;33:617-623.

\bibitem{Tamagnini1997}
Tamagnini P., Troshina O., Oxelfelt F., Salema R., Lindblad P.. Hydrogenases in
  \textit{{N}ostoc} sp. strain PCC 73102, a strain lacking a bidirectional
  enzyme.  {\it Appl. Environ. Microbiol.. } 1997;63:1801-1807.

\bibitem{Axelsson2002}
Axelsson R., Lindblad P.. Transcriptional regulation of \textit{{N}ostoc}
  hydrogenases: {E}ffects of oxygen, hydrogen, and nickel.  {\it Appl. Environ.
  Microbiol.. } 2002;68:444-447.

\bibitem{Eberwine1992}
Eberwine J., Yeh H., Miyashiro K., et al. Analysis of gene expression in single
  live neurons.  {\it Proc. Natl. Acad. Sci. USA. } 1992;89:3010-3014.

\bibitem{Maniatis1982}
Maniatis T., Fritch E.F., Sambrook J.. Synthesis and cloning of cDNA.  in {\it
  Molecular cloning}:212-246Cold Spring Harbor Laboratory 1982.

\bibitem{Lockhart1996}
Lockhart D.~J., Dong H., Byrne M.~C., et al. Expression monitoring by
  hybridization to high-density oligonucleotide arrays.  {\it Nat. Biotechnol..
  } 1996;14:1675-1680.

\bibitem{Singh-Gasson1999}
Singh-Gasson S., Green R.~D., Yue Y., et al. Maskless fabrication of
  light-directed oligonucleotide microarrays using a digital micromirror array.
   {\it Nat. Biotechnol.. } 1999;17:974-978.

\bibitem{Beier2000}
Beier M., Hoheisel J.~D.. Production by quantitative photolithographic
  synthesis of individually quality checked {DNA} microarrays.  {\it Nucleic
  Acids Res.. } 2000;28:E11.

\bibitem{Hasan1997}
Hasan A., Stengele K.-P., Giegrich H., et al. Photolabile protecting groups for
  nucleotides: synthesis and photodeprotection rates.  {\it Tetrahedron. }
  1997;53:4247-4264.

\bibitem{Zhou2002}
Zhou Y., Abagyan R.. Match-only integral distribution ({MOID}) algorithm for
  high-density oligonucleotide array analysis.  {\it BMC Bioinformatics. }
  2002;3:3.

\bibitem{Ermolaeva2001}
Ermolaeva M.~D., White O., Salzberg S.~L.. Prediction of operons in microbial
  genomes.  {\it Nucleic Acids Res.. } 2001;29:1216-1221.

\bibitem{Hofacker1994}
Hofacker I.~L., Fontana W., Bonhoeffer S., Stadler P.~F.. Fast folding and
  comparison of RNA secondary structures.  {\it Monatshefte f\"ur Chemie. }
  1994;125:167-188.

\bibitem{Kaneko2001}
Kaneko T., Nakamura Y., Wolk C.~P., et al. Complete genomic sequence of the
  filamentous nitrogen-fixing cyanobacterium \textit{{A}nabaena} sp. strain
  {PCC} 7120.  {\it DNA Res.. } 2001;8:205-13; 227-53.

\bibitem{Kawashima2003}
Kawashima S., Katayama T., Sato Y., Kanehisa M.. {KEGG API}: {A} Web Service
  Using {S}{O}{A}{P}/{W}{S}{D}{L} to Access the {K}{E}{G}{G} System.  {\it
  Genome Informatics. } 2003;14:673-674.

\bibitem{Baudouin-Cornu2001}
Baudouin-Cornu P., Surdin-Kerjan Y., Marli\`ere P., Thomas D.. Molecular
  evolution of protein atomic composition.  {\it Science. } 2001;293:297-300.

\bibitem{Fauchon2002}
Fauchon M., Lagniel G., Aude J.~C., et al. Sulfur sparing in the yeast proteome
  in response to sulfur demand.  {\it Mol. Cell. } 2002;9:713-723.

\bibitem{Vazquez-Bermudez2003}
V\'azquez-Berm\'udez M.~F., Herrero A., Flores E.. Carbon supply and
  2-oxoglutarate effects on expression of nitrate reductase and
  nitrogen-regulated genes in \textit{{S}ynechococcus} sp. strain {PCC} 7942.
  {\it FEMS Microbiol. Lett.. } 2003;221:155-159.

\bibitem{Vazquez-Bermudez2002}
V\'azquez-Berm\'udez M.~F., Herrero A., Flores E.. 2-{O}xoglutarate increases
  the binding affinity of the {N}tc{A} (nitrogen control) transcription factor
  for the \textit{{S}ynechococcus} \textit{gln{A}} promoter.  {\it FEBS Lett..
  } 2002;512:71-74.

\bibitem{Muro-Pastor1999}
Muro-Pastor A.~M., Valladares A., Flores E., Herrero A.. The \textit{het{C}}
  gene is a direct target of the {N}tc{A} transcriptional regulator in
  cyanobacterial heterocyst development.  {\it J. Bacteriol.. }
  1999;181:6664-6669.

\bibitem{Tumer1983}
Tumer N.~E., Robinson S.~J., Haselkorn R.. Different promoters for the
  \textit{{A}nabaena} glutamine synthetase gene during growth using molecular
  or fixed nitrogen.  {\it Nature. } 1983;306:337-342.

\bibitem{Valladares2003}
Valladares A., Herrero A., Pils D., Schmetterer G., Flores E.. Cytochrome c
  oxidase genes required for nitrogenase activity and diazotrophic growth in
  \textit{{A}nabaena} sp. {PCC} 7120.  {\it Mol. Microbiol.. }
  2003;47:1239-1249.

\bibitem{Jones2002}
Jones K.~M., Haselkorn R.. Newly identified cytochrome c oxidase operon in the
  nitrogen-fixing cyanobacterium \textit{{A}nabaena} sp. strain {PCC} 7120
  specifically induced in heterocysts.  {\it J. Bacteriol.. }
  2002;184:2491-2499.

\bibitem{Schmetterer1994}
Schmetterer G.. {\it The Molecular Biology of Cyanobacteria.}ch.~Cyanobacterial
  respiration., :409-435.
\newblock Kluwer Academic Publisher, NL 1994.

\bibitem{Heddad2002}
Heddad M., Adamska I.. The evolution of light stress proteins in photosynthetic
  organisms.  {\it Comp. Funct. Genom.. } 2002;3:504-510.

\bibitem{Montane2000}
Montan\'e M.~H., Kloppstech K.. The family of light-harvesting-related proteins
  ({LHC}s, {ELIP}s, {HLIP}s): was the harvesting of light their primary
  function?  {\it Gene. } 2000;258:1-8.

\end{thebibliography}
  \bibliographystyle{ama}  

\newpage
\section*{Tables}

\singlespacing

\parindent0pt

\textbf{Table 1} -- Analyzed ORFs. Number of ORFs analyzed and present in each metabolic category of \textit{Nostoc} PCC 7120.
\\\\
\begin{small}
\begin{tabular}{clrcrr}
\hline \multicolumn{2}{l}{\textbf{Metabolic Class}} & \multicolumn{4}{c}{\textbf{Number of ORFs analyzed}}\\ \hline
A & Amino acid biosynthesis & 64 & of & 113 & (57\%) \\
B & Biosynthesis of cofactors, prosthetic groups and carriers & 14 & of & 152 & (9\%) \\
C & Cell envelope & 15 & of & 77 & (19\%) \\
D & Cellular processes & 47 & of & 96 & (49\%) \\
E & Central intermediary metabolism & 40 & of & 72 & (56\%) \\
F & DNA-replication, recombination, and repair & 29 & of & 105 & (28\%) \\
G & Energy metabolism & 20 & of & 100 & (20\%) \\
H & Fatty acid, phospholipid and sterol metabolism & 2 & of & 40 & (5\%) \\
I & Hypothetical & 567 & of & 3573 & (16\%) \\
J & Other categories & 204 & of & 678 & (30\%) \\
K & Photosynthesis and respiration & 98 & of & 157 & (62\%) \\
L & Purines, pyrimidines, nucleosides, and nucleotides & 2 & of & 58 & (3\%) \\
M & Regulatory functions & 36 & of & 360 & (10\%) \\
N & Transcription & 31 & of & 41 & (76\%) \\
O & Translation & 13 & of & 200 & (7\%) \\
P & Transport and binding proteins & 67 & of & 313 & (21\%) \\ \hline
Total & & 1249 & of & 6135 & (20\%)\\ \hline
\end{tabular}
\end{small}

\newpage

\textbf{Table 2} -- Highly Expressed ORFs Under Non-Nitrogen Fixing Conditions. Annotated ORFs that show a more than 2-times higher expression level under non-nitrogen fixing than under nitrogen fixing conditions (the values are negative because nitrogen fixation was taken as reference). All ORFs annotated as putative, hypothetical, or unknown have been omitted. The key for metabolic classes can be found in Table 1.
\\\\
\begin{small}
\begin{tabularx}{\linewidth}{ccXlr}
\hline \textbf{ORF} & \textbf{Class} & \textbf{Submetabolism} & \textbf{Annotation} & \textbf{Expr. Diff.}\\ \hline
all0410   & (A)  & Aromatic amino acid family   & tryptophan synthase beta subunit TrpB   & -2.0\\
alr7354   & (B)  & Thioredoxin, glutaredoxin, and glutathione   & glutathione S-transferase   & -5.9\\
all0166   & (E)  & Polysaccharides and glycoproteins   & alpha,alpha-trehalase   & -6.9\\
asl0873   & (J)  & Adaptations and atypical conditions   & CAB/ELIP/HLIP superfamily   & -4.4\\
asr3042   & "  & "  & "  & -11.3\\
asr3043   & "  & "  & "  & -7.9\\
asl0449   & "  & "  & "  & -2.3\\
all0168   & "  & Other   & alpha-amylase   & -9.5\\
all0275   & "  & "  & glycerophosphoryl diester phosphodiesterase   & -3.4\\
all0167   & "  & "  & maltooligosyltrehalose synthase   & -2.3\\
all0875   & "  & "  & probable alpha-glucanotransferase   & -4.4\\
all0412   & "  & "  & putative zinc-binding oxidoreductase   & -2.2\\
alr2405   & (K)  & Soluble electron carriers   & flavodoxin   & -7.6\\
all0258   & "  & "  & plastocyanin precursor; PetE   & -2.3\\
alr0702   & (O)  & Degradation of proteins, peptides, and glycopeptides   & serine proteinase   & -2.8\\
all2674   & (P)  & Protein modification and translation factors   & ferrichrome-iron receptor   & -2.1\\
all0322   & "  & "  & sulfate-binding protein SbpA   & -4.0\\ \hline
\end{tabularx}
\end{small}

\newpage

\textbf{Table 3} -- Photosynthesis and Respiration. Highly expressed ORFs involved in photosynthesis or respiration. Expression levels are at least 30x above background (300).
\\\\
\begin{small}
\begin{tabularx}{\linewidth}{clXrr}
\hline \textbf{ORF} & \textbf{Submetabolism} & \textbf{Annotation} & \textbf{+N} & \textbf{-N}\\ \hline
all0138  & PS II  & CP47 protein  & 13326 & 9658\\
all0259  & PS II  & cytochrome c550  & 9311  & 13906\\
asr3845  & PS II  & cytochrome b559 alpha-subunit  & 12527 & 10674\\
asr3846  & PS II  & cytochrome b559 beta subunit  & 7606 & 10848\\
alr3421  & Cyt.\ b6/f & plastoquinol--plastocyanin reductase, cytochrome b6; PetB  & 16697 & 12112\\
alr3422  & Cyt.\ b6/f & plastoquinol--plastocyanin reductase, apocytochrome subunit 4; PetD  & 9381 & 13981\\
all0109  & PS I  & Subunit III precursor; PsaF  & 19723 & 23286\\
all0107  & PS I  & Subunit XI; PsaL  & 18416 & 25334\\
all4121  & Electron Carriers  & ferredoxin--NADP(+) reductase  & 5205 & 12290\\
all4148  & Electron Carriers  & ferredoxin I  & 10984 & 14541\\
alr0021  & Phycobilisome  & allophycocyanin alpha subunit; ApcA  & 28417 & 24927\\
alr0022  & Phycobilisome  & allophycocyanin beta subunit; ApcB  & 25066 & 19930\\
asr0023  & Phycobilisome  & core linker protein Lc7.8; ApcC  & 13165 & 16575\\
alr0020  & Phycobilisome  & core-membrane linker protein ApcE  & 7285 & 18364\\
alr0529  & Phycobilisome  & phycocyanin alpha chain; CpcA  & 16927 & 18047\\
alr0528  & Phycobilisome  & phycocyanin beta chain; CpcB  & 20430 & 15579\\
alr0530  & Phycobilisome  & phycocyanin-associated rod linker protein CpcC  & 13993 & 11404\\
alr0532  & Phycobilisome  & phycocyanobilin lyase alpha subunit; CpcE  & 11497 & 18861\\
alr0524  & Phycobilisome  & phycoerythrocyanin alpha chain; PecA  & 11660 & 13985\\
alr0523  & Phycobilisome  & phycoerythrocyanin beta chain; PecB  & 22153 & 16828\\
alr0525  & Phycobilisome  & phycoerythrocyanin-associated rod linker protein PecC  & 4337 & 12756\\
all1842  & NADH DH  & NADH dehydrogenase  & 2762 & 15611\\
all3840  & NADH DH  & Chain J  & 9877 & 14945\\
alr0225  & NADH DH  & Subunit 6; NdhG  & 8989 & 10480\\
alr0226  & NADH DH  & Subunit 4L; NdhE  & 9310 & 12526\\
alr3956  & NADH DH  & Subunit 5  & 7635 & 12323\\
all0006  & ATP Synthase  & Subunit delta; AtpD  & 10175 & 10859\\
all0010  & ATP Synthase  & Subunit a; AtpI  & 15528 & 17550\\
all0011  & ATP Synthase  & Subunit 1; Atp1  & 13805 & 13505\\ \hline
\end{tabularx}
\end{small}

\newpage

\textbf{Table 4} -- Heterocysts. ORFs included in the present study that are specific to heterocysts and show significant expression differences. Expr: expression difference between non-nitrogen and nitrogen fixing condition.
\\\\
\begin{small}
\begin{tabular}{clr}
\hline \textbf{ORF} & \textbf{Description} & \textbf{Expr} \\ \hline
all0521  & two-component response regulator, heterocyst pattern formation protein PatA  & 2.4\\
all0813  & heterocyst-specific glycolipids-directing protein HglK  & 4.6\\
all1430  & heterocyst ferredoxin; FdxH  & 12.9\\
all1431  & HesB protein  & 9.7\\
all1432  & HesA protein  & 8.6\\
all1730  & similar to HetF protein  & 3.4\\
all2512  & transcriptional regulator; PatB  & 2.5\\
all5346  & heterocyst specific ABC-transporter, membrane spanning subunit DevC homolog  & 8.9\\
all5347  & heterocyst specific ABC-transporter, membrane fusion protein DevB homolog  & 3.0\\
alr1603  & putative heterocyst to vegetative cell connection protein (fraH)  & 2.4\\
alr2339  & heterocyst differentiation protein HetR  & 2.1\\
alr2817  & heterocyst differentiation protein HetC  & 5.8\\
alr2818  & heterocyst differentiation protein HetP  & 2.6\\
alr2834 & glycosyltransferase; hepC & 18.5\\
alr2835 & heterocyst specific ABC-transporter; hepA & 5.7\\
alr3234  & similar to heterocyst formation protein HetP  & 6.6\\
alr3648  & heterocyst specific ABC-transporter, membrane spanning subunit DevC homolog  & 2.7\\
alr3649  & heterocyst specific ABC-transporter, ATP-binding subunit DevA homolog  & 3.7\\
alr3698  & heterocyst envelope polysaccharide synthesis protein HepB  & 6.5\\
alr3710  & heterocyst specific ABC-transporter, membrane fusion protein DevB  & 3.3\\
alr3711  & heterocyst specific ABC-transporter, membrane spanning subunit DevC  & 12.4\\
alr3712  & heterocyst specific ABC-transporter, ATP-binding subunit DevA  & 13.1\\
alr4281  & heterocyst specific ABC-transporter, membrane spanning subunit DevC homolog  & 3.5\\
alr4812  & heterocyst differentiation related protein PatN  & 2.3\\
alr5355  & heterocyst glycolipid synthase; HglC  & 4.1\\
alr5358  & ketoacyl reductase; HetN  & 4.0\\ \hline
\end{tabular}
\end{small}

\newpage

\textbf{Table 5} -- Respiratory Terminal Oxidases. Expression differences for respiratory terminal oxidases under nitrogen fixing conditions.
\\\\
\begin{small}
\begin{tabular}{clr}
\hline \textbf{ORF} & \textbf{Description} & \textbf{Expr} \\ \hline
\\
C O X 1 &  & \\ \hline
alr0950 & cytochrome c oxidase subunit II (\textit{coxB}) & less than \textpm2\\
alr0951 & cytochrome c oxidase subunit I (\textit{coxA}) & less than \textpm2\\
alr0952 & cytochrome c oxidase subunit III (\textit{coxC}) & less than \textpm2\\
 &  & \\
C O X 2 &  & \\ \hline
alr2514 & cytochrome c oxidase subunit II (\textit{coxB}) & 4.7\\
alr2515 & cytochrome c oxidase subunit I (\textit{coxA}) & 7.0\\
alr2516 & cytochrome c oxidase subunit III (\textit{coxC}) & 31.8\\
 &  & \\
C O X 3 &  & \\ \hline
alr2731 & cytochrome c oxidase subunit II (\textit{coxB}) & less than \textpm2\\
alr2732 & cytochrome c oxidase subunit I (\textit{coxA}) & 3.3\\
alr2734 & cytochrome c oxidase subunit III (\textit{coxC}) & 4.7\\ \hline
\end{tabular}
\end{small}

\newpage

\textbf{Table 6} -- HLIP-family. Expression differences for the HLIP-family involved in adaptations and atypical conditions. Negative values indicate stronger expression under non-nitrogen fixing conditions.
\\\\
\begin{small}\begin{tabular}{clr}
\hline \textbf{ORF} & \textbf{Description} & \textbf{Expr-Diff} \\ \hline
asl0449  & CAB/ELIP/HLIP-related protein  & -2.3\\
asl0514  & CAB/ELIP/HLIP-superfamily  & less than \textpm2\\
asl0873  & CAB/ELIP/HLIP-superfamily  & -4.4\\
asl2354  & CAB/ELIP/HLIP-related protein  & less than \textpm2\\
asr3042  & CAB/ELIP/HLIP-superfamily of proteins  & -11.3\\
asr3043  & CAB/ELIP/HLIP-superfamily of proteins  & -7.9\\
asl3726  & CAB/ELIP/HLIP-superfamily  & less than \textpm2\\
asr5262  & CAB/ELIP/HLIP-superfamily of protein  & less than \textpm2
\\ \hline
\end{tabular}\end{small}

\section*{Figures}
\begin{minipage}[t][15cm][c]{.9\linewidth}
\begin{figure}[H] 
\begin{center}
\includegraphics[width=9cm]{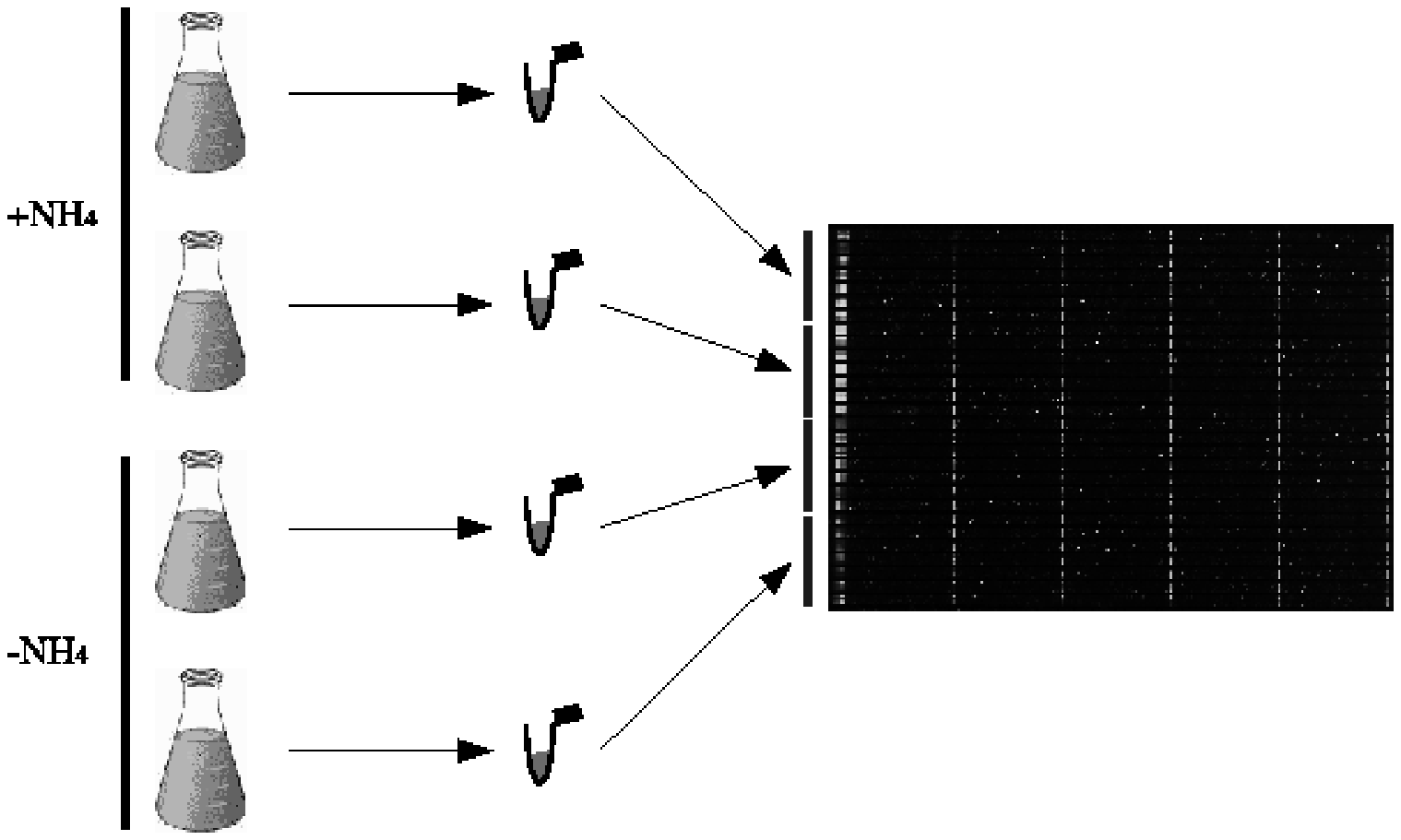}
\end{center}
\textbf{Figure 1} -- Experimental Setup. \textit{Nostoc} PCC 7120 batch cultures were grown under nitrogen fixing (-NH$_4$) or non-nitrogen fixing conditions (+NH$_4$), harvested and subjected to total-RNA purification. After removal of rRNA by means of affinity chromatography, labeled and fragmented cRNA was hybridized to four individual arrays at single physical slides.
\end{figure}
\end{minipage}

\begin{minipage}[t][15cm][c]{.9\linewidth}
\begin{figure}[H] 
\begin{center}
\includegraphics[height=9cm]{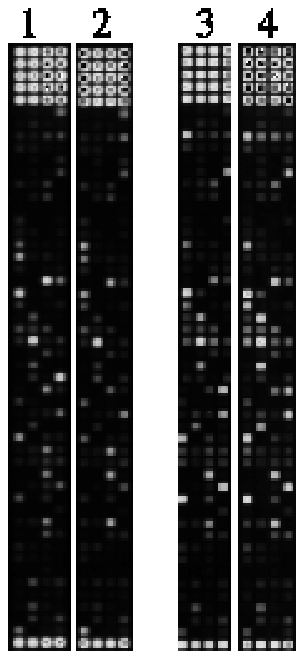}
\end{center}
\textbf{Figure 2} -- DNA-Microarrays. Sections of four arrays used in the present analysis. The bright spots at the top and button of the arrays are controls used by the spot-finding software. (1) non-nitrogen fixing culture (2) non-nitrogen fixing culture, biological replicate (3) nitrogen fixing culture (4) nitrogen fixing culture, biological replicate.
\end{figure}
\end{minipage}

\begin{minipage}[t][15cm][c]{.9\linewidth}
\begin{figure}[H] 
\begin{center}
\includegraphics[width=9cm]{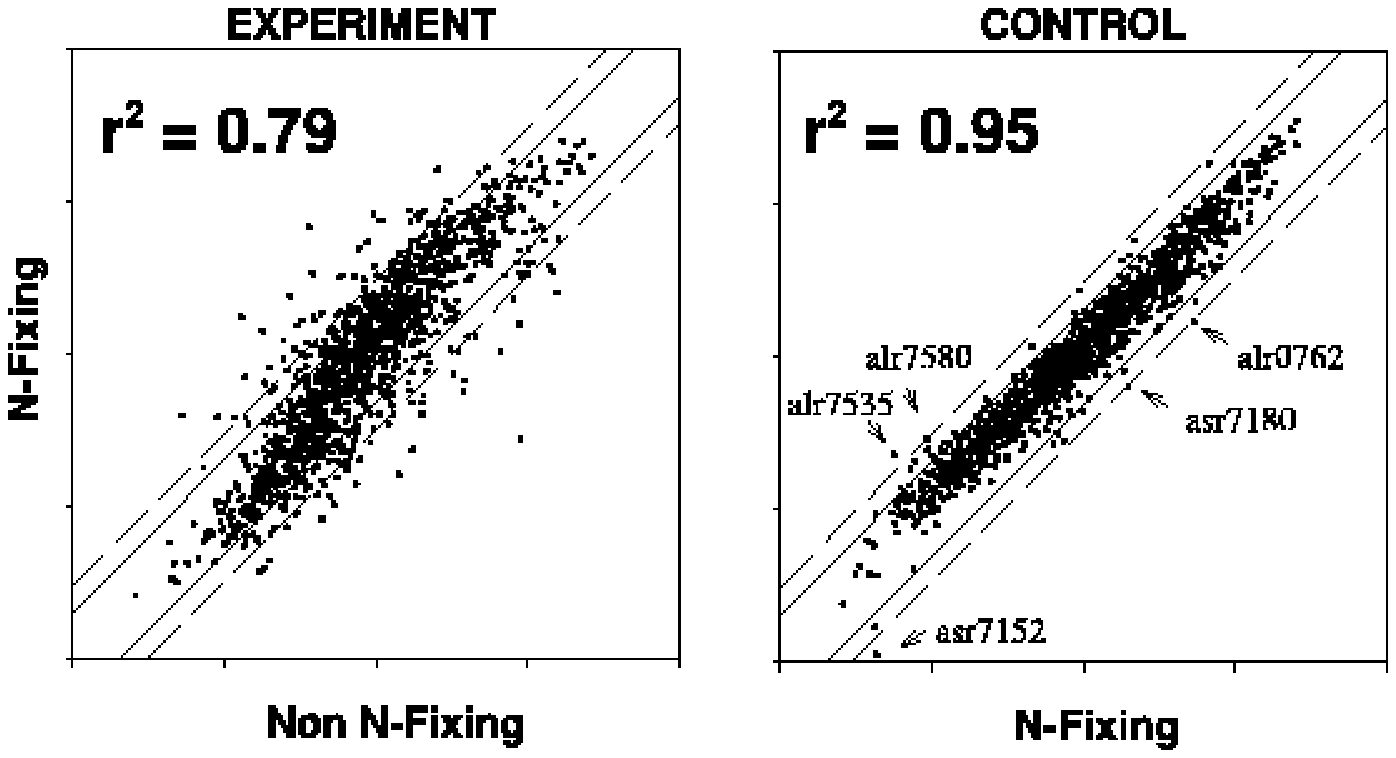}
\end{center}
\textbf{Figure 3} -- Data Quality. For the double-logarithmic plots non-normalized raw data were used. The left plot shows the gene-expression differences between nitrogen fixing (ordinate) and non-nitrogen fixing, ammonia grown (abscissa) cultures of \textit{Nostoc} PCC 7120. The right plot visualizes the gene-expression differences between two individual nitrogen fixing cultures. Significant outliers are marked by arrows. The straight and dashed lines represent 2- and 3-fold expression differences, respectively.
\end{figure}
\end{minipage}

\begin{minipage}[t][15cm][c]{.9\linewidth}
\begin{figure}[H] 
\begin{center}
\includegraphics[width=9cm]{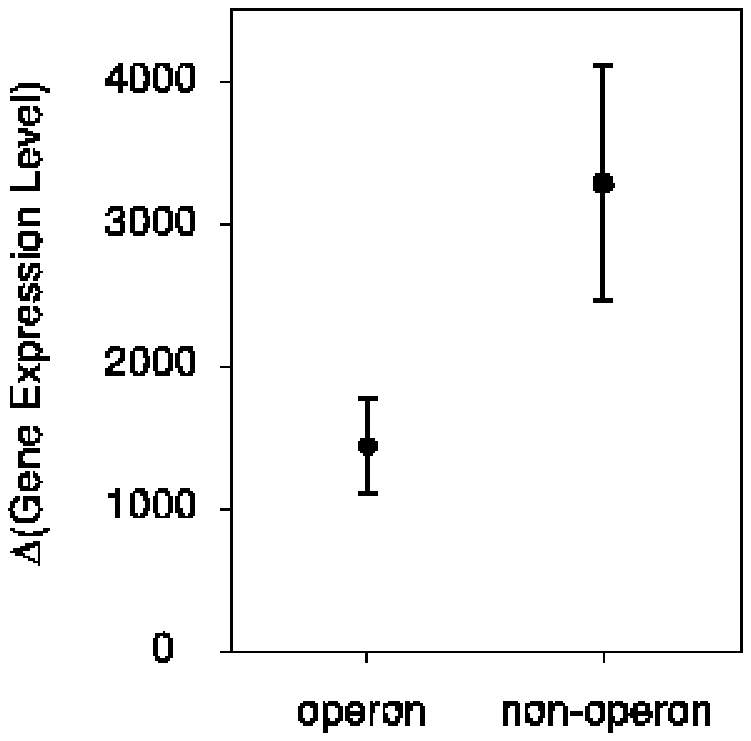}
\end{center}
\textbf{Figure 4} -- Operon Analysis. Depending on whether pairs of ORFs are belonging to a single operon or not, they were split into two groups. Then, the expression difference of each ORF pair was calculated. The mean of the resulting differences in both groups, along with two-times the standard error is plotted. Non-parametric significance tests yield p-values below 0.001.
\end{figure}
\end{minipage}

\begin{minipage}[t][15cm][c]{.9\linewidth}
\begin{figure}[H] 
\begin{center}
\includegraphics[width=9cm]{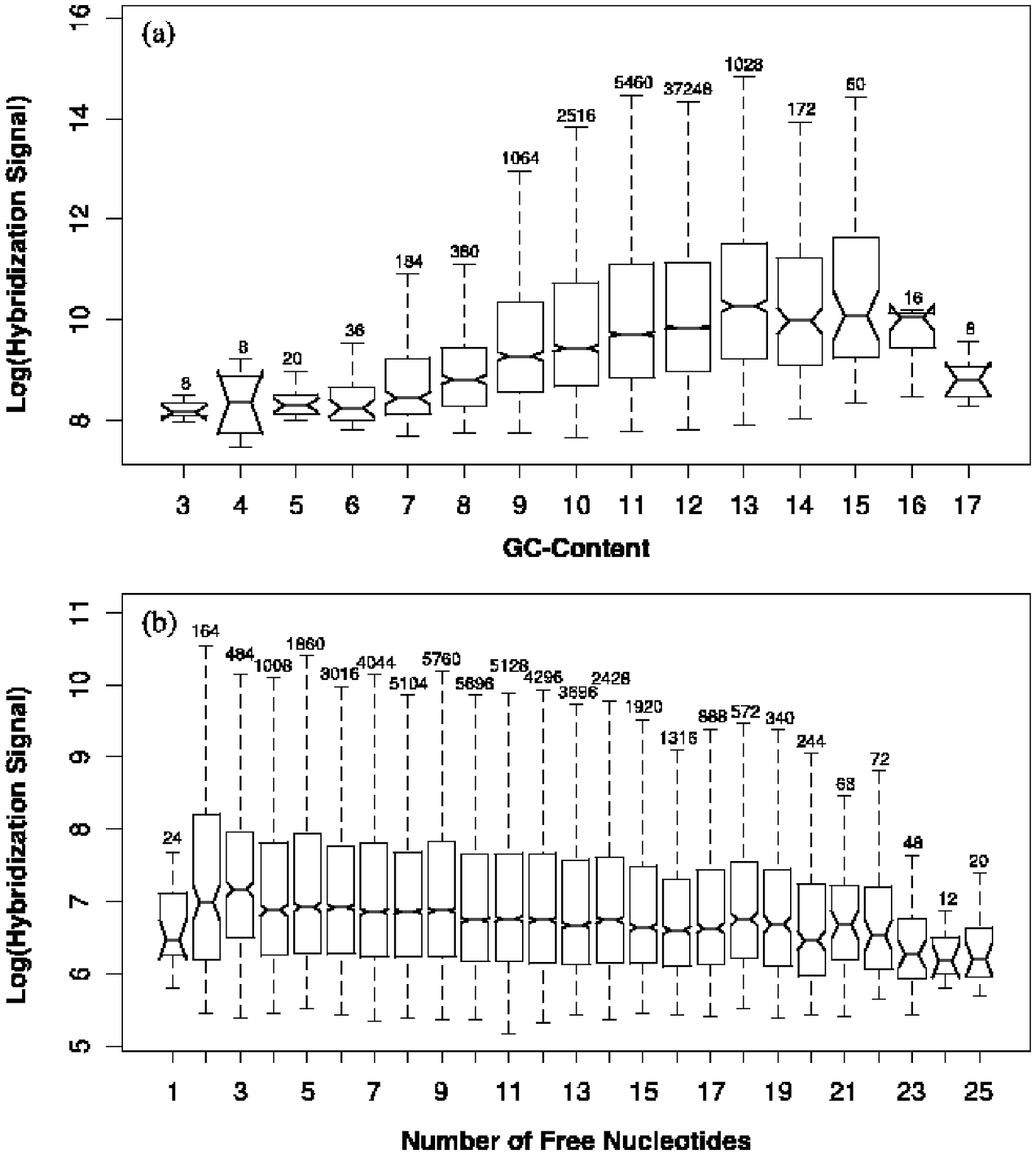}
\end{center}
\textbf{Figure 5} -- Hybridization Effects. The logarithmized intensity of the hybridization signal is plotted (A) against the GC-content of the 25-mere probe sequence, and (B) against the number free secondary structure features (bulges and loops) in the target sequence. The notches around the medians indicate the 95\% confidence interval that the median from one box differs from the median of another box, i.e., if the notches do not overlap the corresponding medians are significantly different. The upper and lower box indicate the second and third quartile, respectively. The plot whiskers extending out from the box to the extreme values. The number above the upper whisker state the number of observations for the corresponding box.
\end{figure}
\end{minipage}

\begin{minipage}[t][15cm][c]{.9\linewidth}
\begin{figure}[H] 
\begin{center}
\includegraphics[width=9cm]{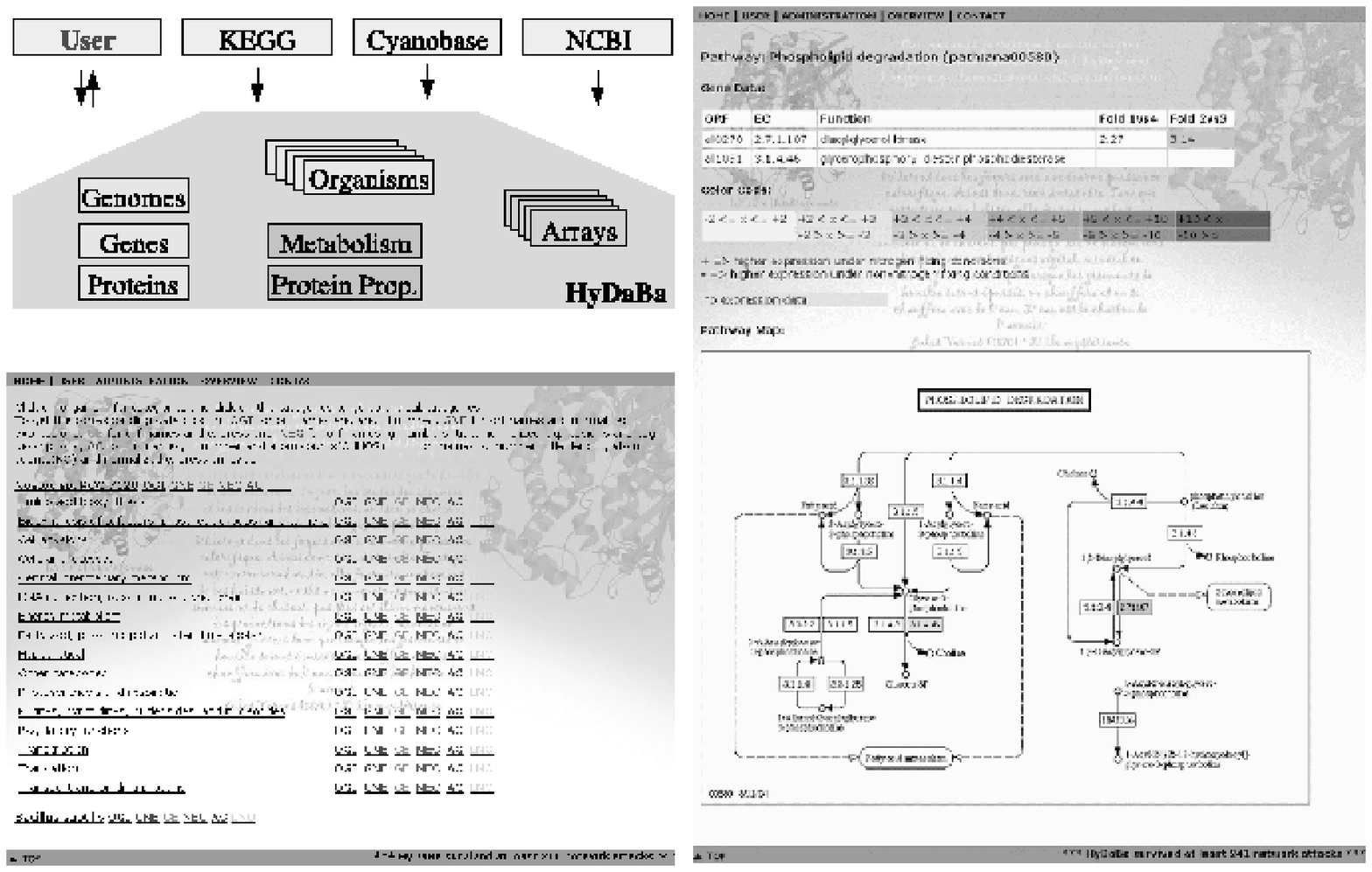}
\end{center}
\textbf{Figure 6} -- HyDaBa Database. All experimental gene-expression data are saved in a relational database (\url{http://hydaba.uni-koeln.de}). This database allows cross-linking of the gene-expression data with genome data from NCBI and Cyanobase and metabolic charts available from KEGG. HyDaBa is based on a Apache Webserver, a MySQL database and a front-end programmed in PHP. All data are publicly accessible via this web interface. Upper left: database outline; Lower left: screenshot of metabolic categories; Right: screenshot of expression data overlayed on metabolic charts.
\end{figure}
\end{minipage}

\begin{minipage}[t][15cm][c]{.9\linewidth}
\begin{figure}[H] 
\begin{center}
\includegraphics[width=9cm]{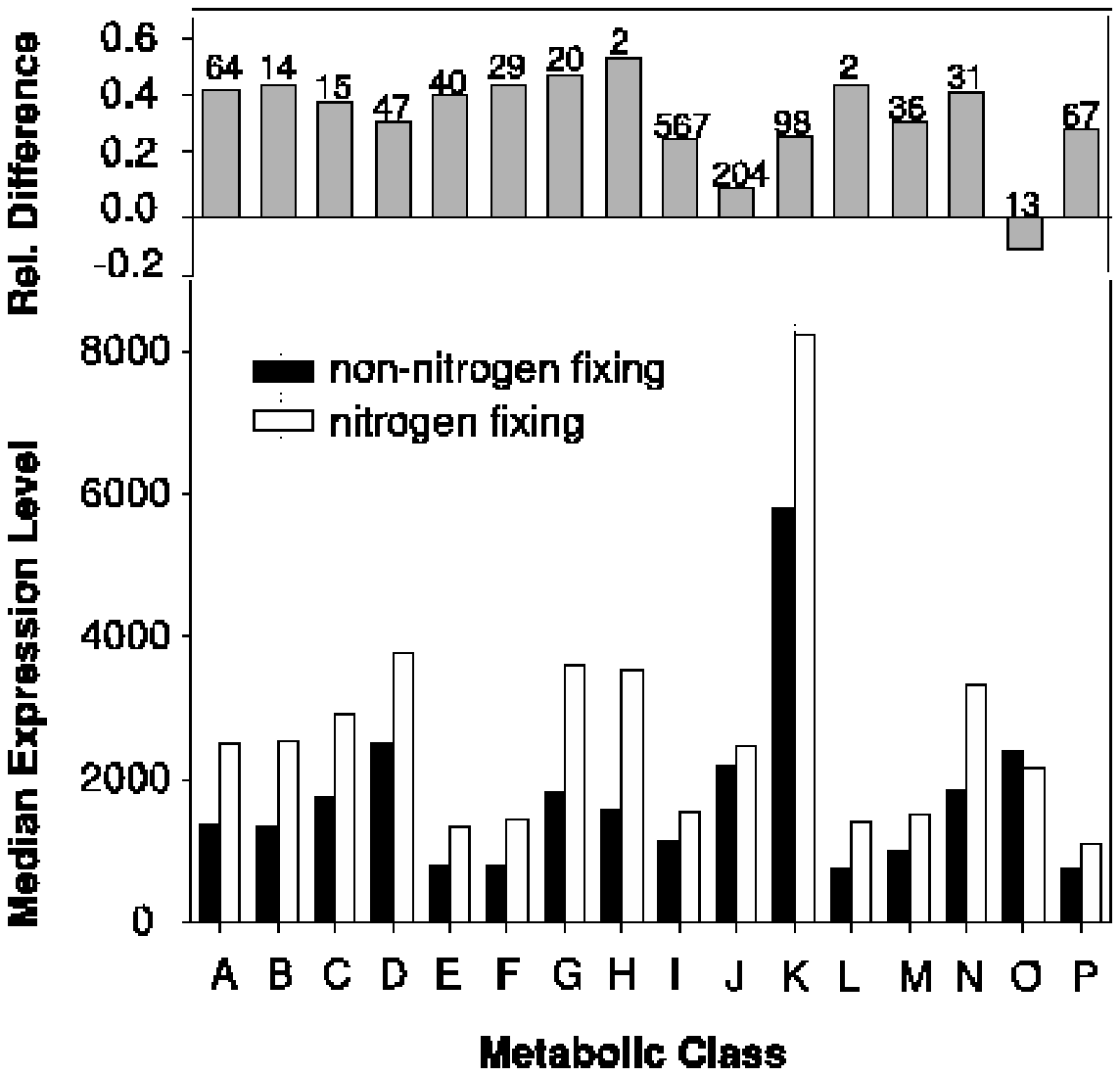}
\end{center}
\textbf{Figure 7} -- Global Expression Differences. Measured gene-expression differences in major metabolic categories. The lower plot shows the median expression level of the analyzed ORFs in the corresponding metabolic class. The upper plot shows the corresponding relative change (maximum range: -1 to 1). The numbers above the bars represent the number of ORFs analyzed in the corresponding category. The legend to the letters denoting metabolic classes can be found in Table 1.
\end{figure}
\end{minipage}

\begin{minipage}[t][15cm][c]{.9\linewidth}
\begin{figure}[H] 
\begin{center}
\includegraphics[width=9cm]{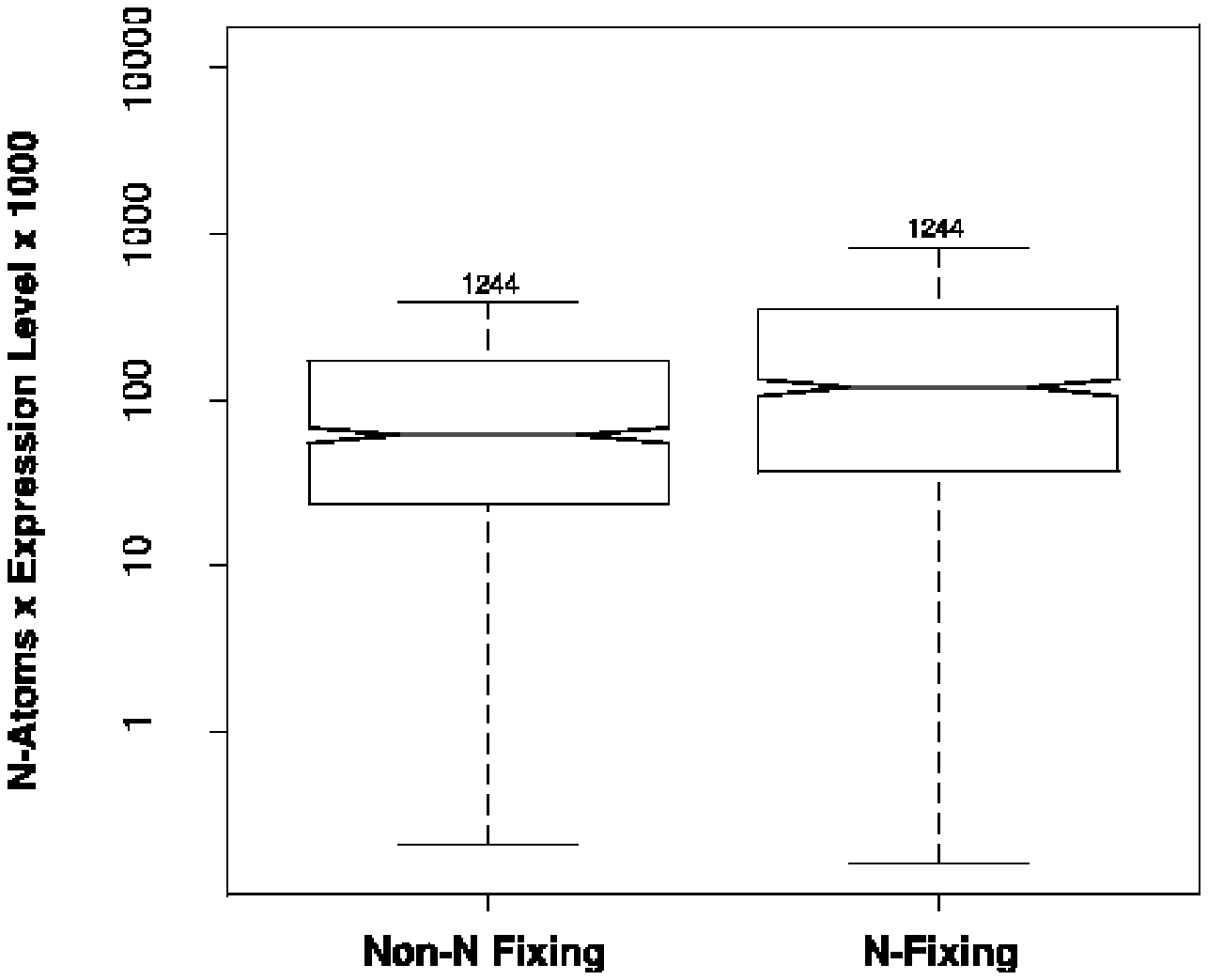}
\end{center}
\textbf{Figure 8} -- Nitrogen Content. Difference of absolute numbers of expressed nitrogen atoms between nitrogen fixing and non-nitrogen fixing conditions. The notches around the median indicate the 95\% confidence interval that the median from one box differs from the median of another box, i.e., if the notches do not overlap the corresponding medians are significantly different. The upper and lower box indicate the second and third quartile, respectively. The plot whiskers extending out from the box to the extreme values. The number above the upper whisker state the number of observations for the corresponding box.
\end{figure}
\end{minipage}

\begin{minipage}[t][15cm][c]{.9\linewidth}
\begin{figure}[H] 
\begin{center}
\includegraphics[height=9cm]{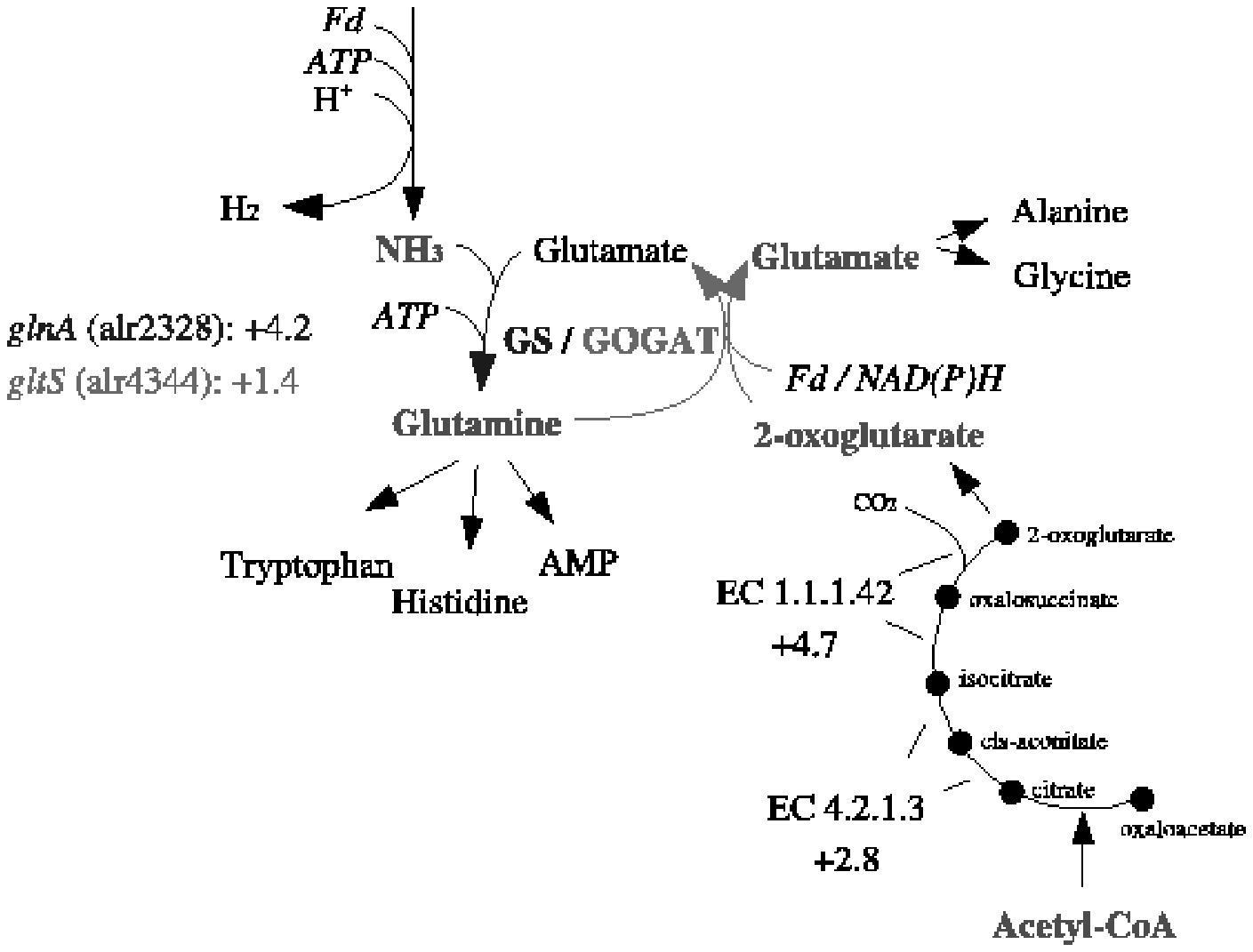}
\end{center}
\textbf{Figure 9} -- Nitrogen Metabolism. The main components involved in nitrogen fixation and assimilation include the nitrogenase (\textit{nif}-genes), the glutamine synthase (GS, \textit{glnA}), and glutamine 2-oxoglutarate amidotransferase (GOGAT, \textit{gltS}). 2-oxoglutarate is synthezised from oxaloacetate and acetyl-CoA by the incomplete TCA-cycle, of which the aconitate hydratase (EC 4.2.1.3) and isocitrate dehydrogenase (EC 1.1.1.42) are shown.
\end{figure}
\end{minipage}

\newpage

\end{document}